# Broadband optical power limiter metalens

Liyi Hsu and Abdoulaye Ndao

*Department of Electrical and Computer Engineering & Photonics Center,
Boston University, 8 Saint Mary's Street, Boston, Massachusetts 02215, USA*
*andao@bu.edu*

**Abstract:**

In recent years, the need for high-power laser is of great interest for different applications ranging from direct-laser processing, light detection, medicine, and lighting. However, high-power lasers with high intensities give rise to fundamental problems for optical detectors and imaging systems with low threshold damage, which still need reliable solutions. Here, we report and numerically demonstrate a hybrid system that synergically combines a broadband optical power limiter with a transmittance difference between on-state (70°C) and off-state (25°C) about 62.5%, and a diffraction-limited broadband metalens from 1534 nm to 1664 nm. Such a metalens power limiter could be used in any system requiring an intermediate focal plane in the optical path to the detector from damage by exposure to high-intensity lasers.

## Introduction

In 1960, after the demonstration of the first laser by Theodore H. Maiman [1], the demand for high-power laser has been tremendously increasing day to day due to the number of applications requiring broadband and high output power. On the other hand, optical detectors and imaging systems are suffering from low threshold damage. To close this gap and overcome such limitations, optical power limiter (OPL) was proposed as an alternative solution [2-4]. An optical power limiter is a key element of protection for sensitive optical sensors and components such as cameras, electronics, and the human eye from high-level laser radiation or power, irradiance, energy transmitted by an optical system below the maximum value, despite the magnitude of the input power. To satisfy such high demand, various methods based on passive optical power limiter have been proposed and demonstrated including nonlinear optical mechanisms which include excited saturation absorption (ESA), two-photon absorption

(TPA), non-linear refraction, induced scattering, and photorefraction [5-12]. Moreover, methods based on multilayer structure or phase change material have been proposed and realized in different platforms [13-16]. So far, proposed methods are either narrowband, single function or require very high input power due to the weak optical nonlinearity. To meet all these demanding specifications of an ideal optical power limiter (OPL), conventional optical limiter often uses an appropriate combination of multiple limiting elements by cascading them in single geometry or using a negative thermal lens, which increases the angle of divergence [17-18]. In addition, most high-power applications require lenses either for processing or for imaging. However, one of the main prominent artifacts in high-power laser applications is thermal lensing [17], producing a focus shift that limits their applications, such as in solid states bulk lasers or amplifiers. Although these methods can overcome low activating threshold, broadband power limiter but are bulky, expensive, or suffer from thermal lensing effects. Despite these proposed methods, a hybrid optical element that synergically combines diffraction-limited broadband metalens and broadband power limiter has remained elusive.

Here, we report and numerically demonstrate a metalens power limiter (MPL) combining a diffraction-limited broadband metalens and an optical power limiter. This platform paves a new way for applications requiring high-power lasers.

To achieve the metalens power limiter (MPL), we use slot waveguide as building blocks. The slot waveguide structure is based on a low-refractive-index slot (Vanadium dioxide) formed between two high-refractive-index (Silicon) waveguides. Vanadium dioxide ($VO_2$) exhibits a metal-insulator phase transition from semiconductor to metal when the temperature increases from room temperature 25°C to 67°C, which has been studied intensely by many researchers. The advantage of using slot waveguides as unit cells is to be able to integrate and overlap dynamic Metal–insulator transitions (MIT) ($VO_2$) and the electric field in the slot to activate the switching function of the devices (from semiconductor to metal). Figure 1 shows the schematic of the proposed hybrid broadband power limiter metalens and its unit-cell. The left metalens shows the diffraction-limited lens function at 25°C and the right one shows a hybrid function combining diffraction-limited lens and OPL at 70°C in which the focusing intensity decreases about 62.5%. The inset at the middle is the unit-cell consisting of two silicon (Si) ridges separated by vanadium dioxide material. The period of the unit-cell (P) is 830 nm, the height of the unit-cell (h) is 880 nm, the gap (g) is 100 nm, the thickness of $VO_2$ is 120 nm, and the metasurface is made of silicon. The design parameters are the width of the cross (W=100 nm), the length of the cross (L=460 nm), and the thickness of the vanadium dioxide.

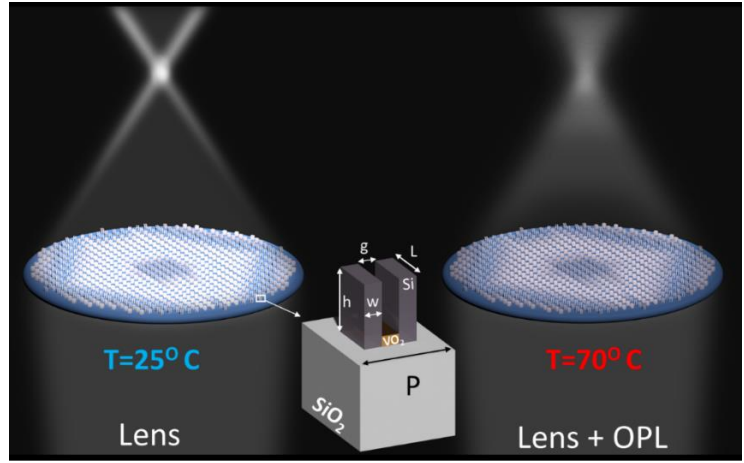

**Fig. 1.** Schematic of the metalens power limiter (MPL) and its unit-cell. The left metalens shows the diffraction-limited lens function at 25°C and the right one shows the MPL combining diffraction-limited lens and OPL at 70°C where the focusing intensity decreases about 62.5%. The inset at the middle is the unit-cell consisting of two silicon ridges separated with vanadium dioxide material in the gap. The period of the unit-cell (P) is 830 nm, the height of the unit-cell (h) is 880 nm, the gap (g) is 100 nm, the thickness of $VO_2$ is 120 nm, and the metasurface is made of silicon. The design parameters are the width of the cross (W=100 nm), the length of the cross (L=460 nm), and the thickness of the vanadium dioxide.

### 1. Design strategy of the MPL

To design the hybrid system, we use a slot waveguide as a building block in view to increase the interaction between the electric field of the incident wave and the $VO_2$ layer and to provide a broadband response. The slot waveguide consists of two high-index ridges (Si), separated by a narrow low-index gap ($VO_2$). Due to such refractive index discontinuity, this structure allows one of the propagating modes to confine its energy within the slot region [19].

The advantage of using slot waveguides as unit cells is twofold. Firstly, unlike resonant modes, which are intrinsically narrow band, waveguide structure provides propagating modes which enable broadband response [20-23]. The second one consists of maximizing the confinement and interaction between the strong electric field supported by the slot waveguide and the phase transition material ($VO_2$) to activate the power limiting function. To optimize the design of the power limiter, the maximum overlap of the electric field of the slot waveguide and the vanadium dioxide must be achieved. We performed numerical simulations using finite-difference time-domain (FDTD) solver Lumerical (field distributions for different configurations) presented in Fig. 2. The structure is illuminated at normal incidence with circular polarization (CP) light. Figure 2(a) shows the evolution of the design principle in order or to maximize the overlap between the vanadium dioxide and the electric field in the gap. The substrate is silicon dioxide ($SiO_2$) with refractive index 1.53, and the two bars are silicon with refractive index 3.48. We first consider a standard slot waveguide and calculate the electric field in the gap (Fig. 2(a)) which is mostly located on top of the slot, therefore requires to have a similar thickness of $VO_2$ as the silicon for efficient design. Thus, it would make the realization challenging. To increase the electric field at the interface between the substrate and the thickness of $VO_2$, we then embedded the slot waveguide into the substrate (Fig. 2(b)). This embedded slot considerably increases the overlap between $VO_2$ and the E-field (Fig. 2(c)) and minimize its thickness ($VO_2$).

To illustrate the optical power limiting function, figure 2(d) presents the normalized transmittance spectrum of the embedded slot waveguides nanoparticles at different temperatures at T=25°C and T=70°C. The normalized transmittance is about 100% at room temperature (blue) while the normalized transmittance for T=70°C is about 37%. Thus, this large transmittance difference between on-state (70°C) and off-state (25°C) about 62.5%, shows the successful implementation of a highly efficient optical power limiter.

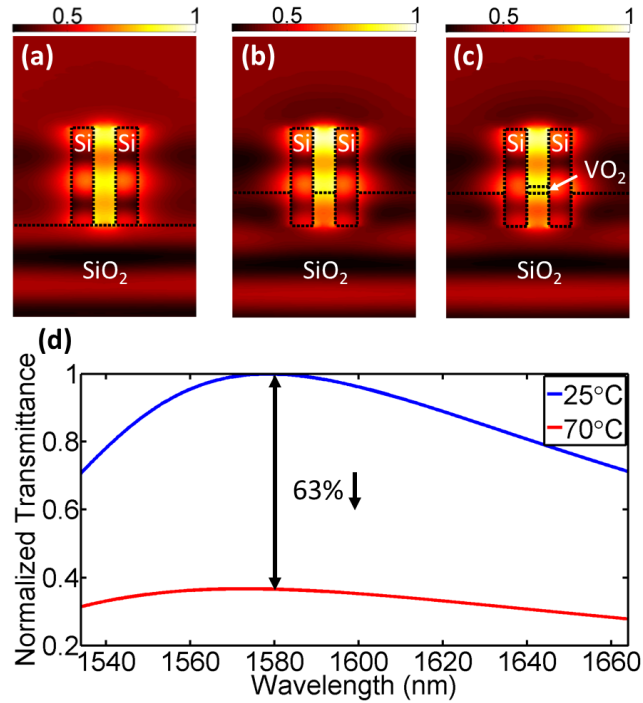

**Fig. 2.** Design principle of MPL. (a)-(c) Electric fields of different unit cells. (a) Electric field of slot waveguide. (b) Electric field of partially embedded slot waveguide. (c) Electric field of partially embedded slot waveguide with VO$_2$ layer in the middle. (d) Normalized transmittance of the unit cell for different temperatures. The blue line represents room temperature transmittance and the red one the temperature at 70ºC. This large transmittance difference between the two states (averagely 62.5%) thus shows the successful implementation of a highly efficient optical power limiter.

## 2. Thermal performance of the broadband MPL

To further investigate the performance of the OPL, we used a finite-difference solver (Lumerical Heat solver) to calculate the mean temperature of the VO$_2$ film. The optical performance is first calculated by Lumerical FDTD and the absorption power is obtained by the equation:

$$P_{abs} = \frac{1}{2}\varepsilon_i \int |E|^2 dV \quad (1)$$

P$_{abs}$ is the absorption power; ε$_i$ is the imaginary part of permittivity; E is the electric field, and its integration region is the volume of VO$_2$ because it is the only lossy material in this wavelength range. The absorption power is then imported to the Heat solver as the heat source. Adiabatic boundary conditions are used in the in-plane direction to mimic an infinite array. The simulated mean temperature of VO$_2$ film as a function of the input power in a steady state is shown in Fig. 3(a).

We can observe that the input intensity needed to heat the VO$_2$ to its phase-transition temperature of 67°C (orange line) is 3.5 kW/cm$^2$. The insets are the temperature distributions at different input intensities. The left one with the blue box is at 3.5 kW/cm$^2$ and the right one with the pink box is 8 kW/cm$^2$. In addition, to steady-state simulation, the transient analysis is also important because, for applications requiring high power limiters, pulsed laser sources are often used. We used pulse bandwidths 10 ns in the paper. The transient simulation result is shown in Fig. 3(b). We can observe that the intensity for achieving the phase-transition temperature is 0.5 MW/cm$^2$ which is significantly higher than the steady-state condition. It is because the absorbed energy is in a much shorter time. In the case of 0.3 MW/cm$^2$, the temperature will increase to 50°C while it can be up to 100°C for 1MW/cm$^2$. The small thermal relaxation time (less than 100 ns) of the OPL at high power, thus protect the optical power limiter from damage.

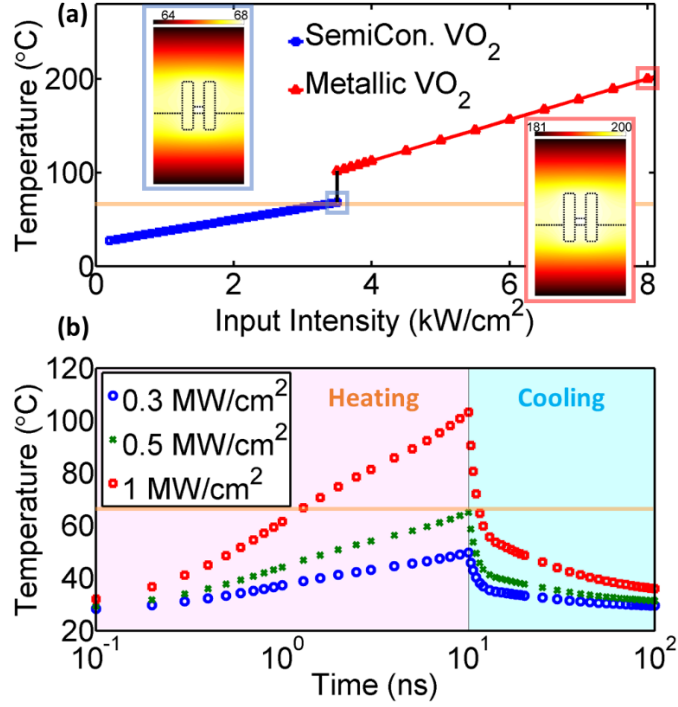

**Fig. 3.** Thermal performance of the MPL. (a) simulated average temperature of the VO$_2$ layer with different input intensity in a steady state. The insets are the temperature distributions at different input intensities. The left one with a blue box is at 3.5 kW/cm$^2$ and the right one with a pink box is 8 kW/cm$^2$. The orange line is the temperature of phase-changing (67°C). (b) Temporal thermal characteristics for 10 ns pulses width of peak intensities of 0.3, 0.5, and 1 MW/cm$^2$. The orange line represents the temperature of phase-changing (67°C). The small thermal relaxation time (less than 100 ns) of the OPL at high power, thus protect the optical power limiter from damage.

### 3. Broadband metalens power limiter (MPL)

To synergically combine the two functions OPL and broadband metalens, we use the same designed unit-cell of the power limiting as a building block. For a normal incidence of CP light, the transmitted electric field (E$^t_{L/R}$) of the unit-cell rotated by an angle θ can be calculated through the Jones matrix as follows [24]:

$$E_{L/R}^t = E_s \hat{e}_{L/R} + E_c e^{\pm i2\theta} \hat{e}_{R/L} \quad (2)$$

The two terms in equation 2 represent respectively same circular polarization of the transmitted wave and its opposite. $\hat{e}_{R/L} = (\hat{e}_x \pm i\hat{e}_y)/\sqrt{2}$ is Jone's vector of LCP or RCP. $E_s = (e^{i\phi/2} + e^{-i\phi/2})/2$ and $E_c = (e^{i\phi/2} - e^{-i\phi/2})/2$. φ represents the phase difference between the x and y directions. When φ is π, the opposite CP wave will provide the phase shift ±2θ (Pancharatnam-Berry phase).

To focus light to a point for a normal incident plane wave, the phase profile of the wavefront as a function of position (r) along the metalens must satisfy the parabolic equation given by the relation [25-30]:

$$\phi(r) = -\frac{2\pi}{\lambda}\left(\sqrt{r^2 + F^2} - F\right) + g \quad (3)$$

where ϕ (r) is the phase profile required, λ is the wavelength, F is the focal length, r is the radial position, and g is a reference phase independent of r. The reference phase can be an arbitrary function of frequency because only the spatial phase difference matters for the interference of waves at the same frequency after their interaction with the lens. Here, we consider the case of g = 0 i.e. the reference phase is at the center of the metalens (r = 0). To demonstrate the hybrid function system, using the same unit cell with different rotation angles, we design a metalens with a diameter of 60 μm and a numerical aperture NA = 0.32.

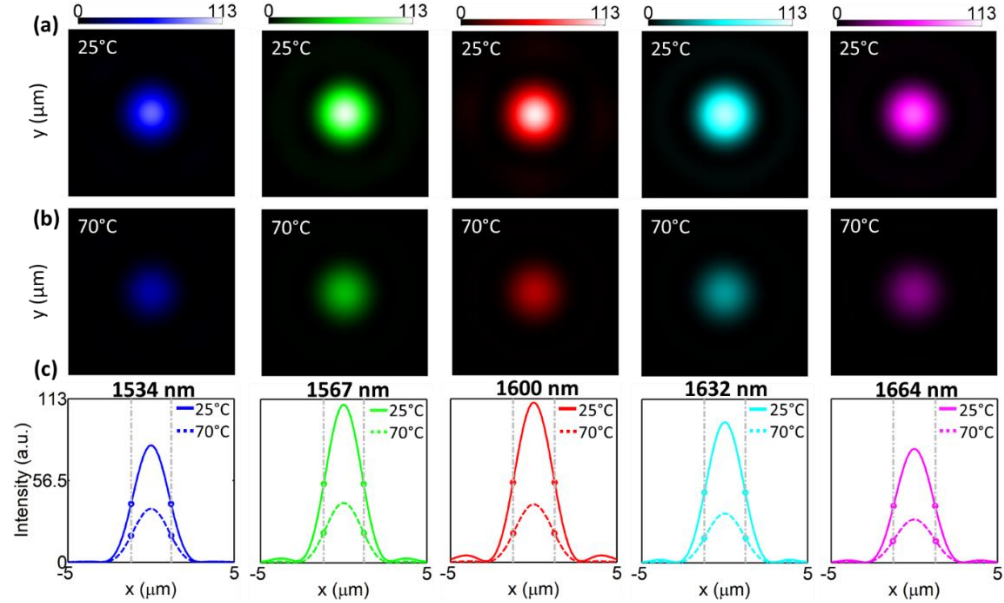

**Fig. 4.** Simulated intensity profiles at the focal plane of the MPL. The numerical aperture NA = 0.32, at different wavelengths from 1534 nm to 1664 nm for (a) 25°C and (b) 70°C. (c) Cross-section of figure 4(a) and 4(b) at different wavelengths from 1534 nm to 1664 nm. The solid lines are at 25°C and the dish lines are at 70°C. The circles and the grey dish lines indicate the position of half maximum.

Figures 4(a), (b) show simulated intensity profiles at the focal plane of the metalens at different wavelengths from 1534 nm to 1664 nm for 25°C and 70°C. One can conclude that the intensities of focal spots decrease more than half from 25°C to 70°C in the design wavelength bandwidth. Figure 4(c) is the cross-section of Fig. 4(a) and Fig. 4(b) at different wavelengths from 1534 nm to 1664 nm. The solid lines are for 25°C, and the dash lines are for 70°C. The circles and the grey dish lines indicate the position of half maximum. We can observe that the full widths at half maximum (FWHM) remain the same while the maximum focus intensities decrease. It confirms that our hybrid metalens can limit the optical power while keeping diffraction-limited focusing. Figure 5(a) shows the maximum intensity at the focal plane of the metalens at different wavelengths from 1534 nm to 1664 nm for 25°C (blue color) and 70°C (red color). At room temperature (25°C), the maximum intensity is observed at 1580 nm. When the temperature increases to 70°C, the maximum intensities reduce averagely by 62.5%. The hybrid metalens result is consistent with the unit cell transmittance in Fig. 2(d) because the coupling difference between unit-cell is ignorable. We also present in Fig. 5(b) the Strehl ratio defined as the ratio of the peak focal spot intensity of the hybrid metalens to the focal spot intensity of an ideal Airy disk. We can observe that the Strehl ratio in the entire bandwidth at different temperatures (25°C in blue color and 70°C in red color) is larger than 0.9. We have thus successfully implemented a MLP combining OPL and a diffraction-limited broadband metalens from 1534 nm to 1664 nm.

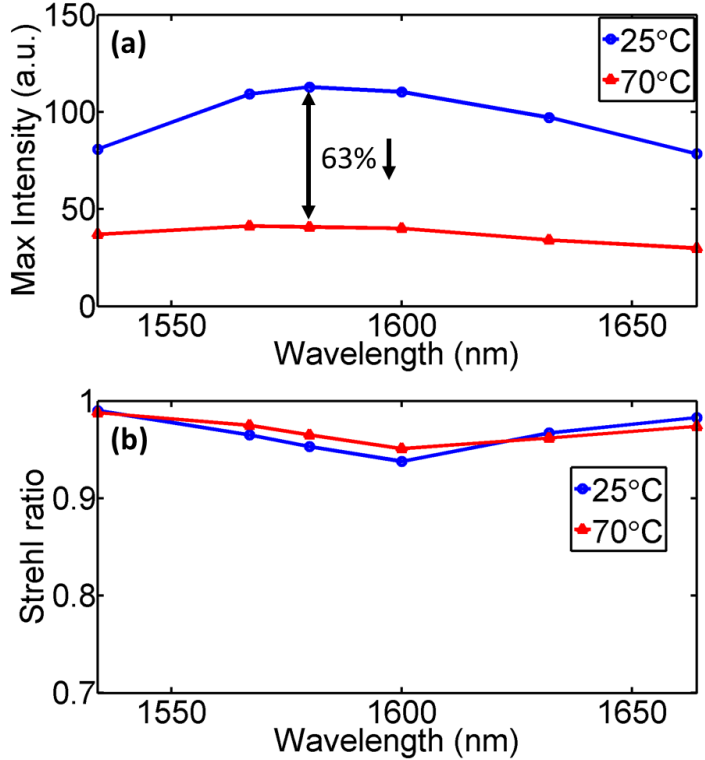

**Fig. 5.** Optical performance of the MPL. (a) Max intensity at the focal plane of the metalens at different wavelengths from 1534 nm to 1664 nm for 25°C in blue color and 70°C in red color. (b) Strehl ratio at different wavelengths from 1534 nm to 1664 nm at 25°C in blue color and 70°C in red.

In conclusion, we have designed and demonstrated a new hybrid system synergically that combines a broadband optical power limiter (OPL) with a transmittance difference between on- state (70°C) and off-state (25°C) about 62.5% and a diffraction-limited broadband metalens from 1534 nm to 1664 nm. The multifunction broadband MPL is achieved by combining a slot waveguide and the $VO_2$ MIT properties. To the best of our knowledge, this is the first time such a hybrid system is proposed and demonstrated. Our proposed hybrid devices could be used in different frequency ranges from optics to microwaves and may find employment either in protective applications or any other systems where high-power limiters and metalens are essential.

## Acknowledgments

The authors gratefully acknowledge the financial support of Boston University start-up funding.